\documentclass[conference]{IEEEtran}
\IEEEoverridecommandlockouts
\usepackage{cite}
\usepackage{amsmath,amssymb,amsfonts}
\usepackage{algorithmic}
\usepackage{graphicx}
\usepackage{textcomp}
\usepackage{xcolor}
\usepackage{subfigure}
\usepackage{colortbl,multirow,hhline}
\usepackage{bbding}
\def\BibTeX{{\rm B\kern-.05em{\sc i\kern-.025em b}\kern-.08em
    T\kern-.1667em\lower.7ex\hbox{E}\kern-.125emX}}
\begin{document}

\title{ADF-GA: Data Flow Criterion Based Test Case Generation for Ethereum Smart Contracts\\
}

\author{\IEEEauthorblockN{Pengcheng Zhang}
\IEEEauthorblockA{\textit{College of Computer and Information} \\
\textit{Hohai University}\\
Nanjing, China \\
pchzhang@hhu.edu.cn}
\and
\IEEEauthorblockN{Jianan Yu}
\IEEEauthorblockA{\textit{College of Computer and Information} \\
\textit{Hohai University}\\
Nanjing, China \\
yu\_poppy@qq.com}
\and
\IEEEauthorblockN{Shunhui Ji\textsuperscript{\Envelope}}
\IEEEauthorblockA{\textit{College of Computer and Information} \\
\textit{Hohai University}\\
Nanjing, China \\
shunhuiji@hhu.edu.cn}
}

\maketitle

\begin{abstract}
Testing is an important technique to
improve the quality of  Ethereum smart contract programs.
However, current work on testing smart contract only focus on static problems of smart contract programs. A data flow oriented test case generation approach for dynamic testing of smart contract programs is still missing. To address this problem, this paper proposes a novel test case generation approach, called ADF-GA (\underline{A}ll-uses \underline{D}ata \underline{F}low  criterion based test case generation using \underline{G}enetic \underline{A}lgorithm),
for \emph{Solidity} based Ethereum smart contract programs.
ADF-GA aims to efficiently generate a valid set of test cases via three stages. First, the corresponding program control flow graph is constructed from the source codes. Second, the generated control flow graph is analyzed to obtain the variable information in the \emph{Solidity} programs, locate the \emph{require} statements, and also get the definition-use pairs to be tested. Finally, a genetic algorithm is used to generate test cases, in which an improved fitness function is proposed to calculate the definition-use pairs coverage of each test case with program instrumentation. Experimental studies are performed on several representative \emph{Solidity} programs. The results show that ADF-GA can effectively generate test cases, achieve better coverage, and reduce the number of iterations in genetic algorithm.
\end{abstract}

\begin{IEEEkeywords}
blockchain, smart contract, \emph{Solidity}, test case generation, genetic algorithm, fitness function
\end{IEEEkeywords}

\section{Introduction}
Blockchain technology is a groundbreaking technology that truly realizes decentralization~\cite{a1}. Ethereum, as one of the mainstream blockchain platforms, attracts a large number of developers and researchers because of its simplicity, convenience and openness to all uses~\cite{a2}. \emph{Solidity} is a main programming language for writing smart contracts on Ethereum, derived from \emph{JavaScript}, \emph{Python}, and \emph{C++}~\cite{b1}.
Recently, it is frequent to find serious security vulnerabilities caused by insecure programming in smart contracts, which cause significant losses. For example, the re-entrancy vulnerability in the DAO~\cite{a3} led to the loss of \$$50$ million worth of Ethers at that time. In addition, due to the special characteristics of blockchain, smart contracts cannot be modified once deployed. Therefore, it is critical to ensure the security and function completeness of a smart contract program before it is deployed. Nevertheless, it needs to constantly improve the sufficient testing of \emph{Solidity} smart contracts, considering the impact of its execution environment and its characteristics on the execution of the smart contract.

In recent years, more and more researchers adopt different methods to develop different tools and make contributions to smart contracts security testing~\cite{b2,b3,b4,b5,b6}.
Static analysis and testing is still the main idea of most of these approaches or tools~\cite{b2,b3} due to the limitations of the blockchain environment. These approaches or tools analyze and detect the bugs or problems in smart contract codes according to the known or predefined error types, and then give the detection results by matching the existing vulnerability one by one. They may also statically reason about a program path-by-path through symbolic execution~\cite{b2}. Other approaches apply a fuzzy test to detect specific vulnerabilities in smart contracts~\cite{b4,b5,b6}. While these approaches make some progress on smart contract testing, the state-of-the-art testing generation approaches for smart contracts still suffer the following two main limitations:
\begin{itemize}
  \item The research on dynamic testing of smart contracts is still insufficient. The existing researches mainly detect known vulnerabilities in contracts in static ways. Unexpected errors that may happen during program execution can not be tested by static approaches. Consequently, a dynamic data flow oriented testing approach for smart contract programs is necessary.
  \item Existing test case generation approaches are only designed for traditional development languages, such as \emph{Java} and \emph{C\#}. Some of them are mature, but cannot deal with the specific features of \emph{Solidity} smart contracts, such as the \emph{require} statements.
 \end{itemize}


To address the aforementioned limitations, we propose a novel all-uses data flow criterion based test case generation approach for \emph{Solidity} smart contracts, 
 called ADF-GA (\underline{A}ll-uses \underline{D}ata \underline{F}low  criterion based test case generation using \underline{G}enetic \underline{A}lgorithm), which means that every $dup$ (definition-use pair)~\cite{b9} needs to be covered in the testing. The key ideas of ADF-GA are described as follows: \emph{1)} The CFG (control flow graph) of a smart contract is constructed as an intermediate representation for the source codes to extract the internal information of the contract.
\emph{2)} Data flow analysis is performed to obtain the information of variables and the $dup$s to be tested. \emph{3)} By constructing a reasonable chromosome coding structure, we apply GA (genetic algorithm)~\cite{b10} to optimize the generation of test cases for smart contracts, in which an improved fitness function for calculating individual fitness value to guide population evolution, which makes ADF-GA  effectively generate test cases that satisfy the criterion.

In summary, the main contributions of this paper are described as follows:
\begin{itemize}
  \item To the best of our knowledge, data flow criterion based test case generation for dynamic testing of smart contracts is proposed for the first time, which can effectively test data information during the execution of the smart contracts.
  \item An improved fitness function is proposed to emphasize the coverage of the \emph{require} statements related $dup$s by setting a weighted parameter, which helps to perform better selection operation of individuals in GA.
  \item Finally, based on several representative \emph{Solidity} programs,
  we design a set of comparative experiments to compare ADF-GA with two other approaches. The experimental results show the effectiveness and efficiency of ADF-GA.
\end{itemize}

The structure of the paper is organized as follows: Section~\ref{sec RelatedWork} surveys state-of-the-art smart contract testing approaches and discusses their limitations. Section~\ref{sec prelimaries} introduces the preliminary knowledge. Section~\ref{sec approach} presents the details of our approach. Section~\ref{sec_validation} performs the experimental studies based on several data sets. Section~\ref{sec_conclusions} concludes the paper and plans our future work.

\section{Related Work}\label{sec RelatedWork}

\subsection{Smart Contracts Vulnerability Detection and Testing}\label{contract test}

Existing approaches and tools can be summarized into two main categories in terms of testing technology: static analysis based vulnerability detection~\cite{b2,b3} and fuzzy test-based vulnerability detection~\cite{b4,b5,b6}.

Static analysis approaches and tools have many different implementations, including static code analysis, symbol execution, and etc. Static analysis can be effective in detecting program vulnerabilities, especially for coding errors and some known error types. Luu et al.~\cite{b2} proposed an approach based on symbolic execution and built a tool to find potential security bugs. This approach is the first work to use CFG of smart contracts for information extraction. Tsankov et al.~\cite{b3} presented another tool to analyze a smart contract program according to the predefined security patterns by transforming the EVM (Ethereum Virtual Machine) bytecode provided as the input into a stackless representation in static-single assignment form and then the corresponding semantic facts are inferred. Tikhomirov et al.~\cite{b11} provided a comprehensive classification of code issues in \emph{Solidity} and implemented an extensible static analysis tool to detect them.

Fuzzy test-based approaches are used to generates fuzzing inputs to detect vulnerabilities. Liu et al.~\cite{b4} presented an analyzer for re-entrancy bugs detection. The tool firstly converts the smart contract to \emph{C++} using an intermediate representation; and then fuzzes the \emph{C++}. Finally, a bug report is given. This analyzer only addresses the re-entrancy attack. Besides, Liu et al.~\cite{b5,b6} proposed a more perfect fuzzy test approach, which is used for fuzzy testing of seven types of vulnerabilities. The approach generates    based on the ABI (Application Binary Interface) specifications of the \emph{Solidity} smart contracts and detect vulnerabilities based on defined test oracles. The approach has a high detection accuracy for predefined vulnerabilities.

However, existing testing approaches are based on already defined vulnerabilities. Furthermore, they mainly use static analysis techniques and do not consider dynamic executions of smart contracts.

\subsection{Data Flow Criterion Based Test Case Generation}\label{test case}

Dynamic test detects bugs through the program execution and the analysis of the program running results, common test methods including logical test, path test, data flow test. Among them, data flow test focus on the interaction of data flow in a program, some studies show that test cases generated based on data flow criteria are better~\cite{b8}.

Rapps et al.~\cite{b12} extended and defined the concept of data flow analysis and proposed different data flow criteria. On this basis, Pande et al.~\cite{b13} implemented data flow analysis for C for test case generation. Harrold et al.~\cite{b14} extended data flow analysis to object-oriented programs and put forward object-oriented programs test based on data flow criterion. Nayak et al.~\cite{b15} applied the particle swarm optimization algorithm to implement data flow testing. Deng et al.~\cite{b16} proposed to combine GA with data flow analysis to generate test cases. Girgis et al.~\cite{b17} proposed a test case generation method based on the all-uses strategy and used it to test \emph{C\#} programs. Vivanti et al.~\cite{b18} applied the GA to the data flow test of object-oriented programs and proved the validity of test case generation methods based on data flow criterion in practice.

Although there are many researches on data flow based test case generation, they are designed for traditional languages such as
\emph{C++} and \emph{Java}. Due to the unique structure, statements as well as the use of variables in \emph{Solidity}, these approaches cannot be directly used.
\begin{figure}[t]
\centerline{\includegraphics[width=8.6cm,height=3cm]{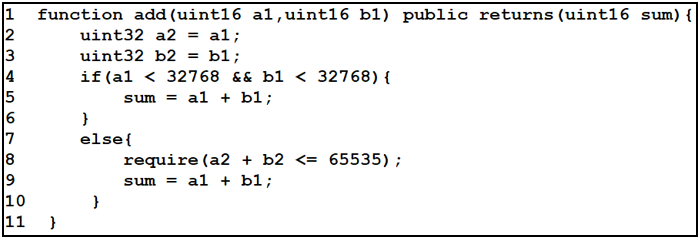}}
\caption{Sample program: function \emph{add}}
\label{fig1}
\end{figure}

\begin{figure*}[t]
\centering
   \includegraphics[width=0.8\linewidth]{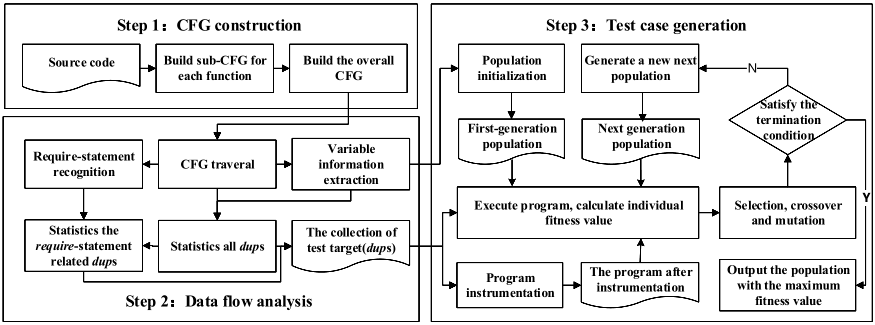}
      \caption{The main architecture of ADF-GA} \label{fig3}
\end{figure*}

\section{Background}\label{sec prelimaries}

\subsection{Smart contracts in Solidity}\label{SCS}
Smart contracts are specific programs, which run on blockchain platforms (e.g., Ethereum). Most Ethereum smart contracts are written in \emph{Solidity}, whose programming style is similar to \emph{JavaScript} and \emph{Python}. Smart contracts written in \emph{Solidity} are specifically deployed and run on the Ethereum in the EVM bytecode format~\cite{b19}.

There are three main structures in smart contract programs: \emph{sequential}, \emph{selection}, and \emph{loop} structure~\cite{b19}. \emph{Solidity} has some different attributes with traditional development languages. Fig.~\ref{fig1} shows a function ``\emph{add}'' written in \emph{Solidity}, which can get the sum of two variables of type \emph{uint16}. The \emph{require} statement is used at line 8 to qualify a program execution condition, and the program continues execution only if this \emph{require} statement condition is satisfied.

In addition, in \emph{Solidity}, the type of numeric variable is relatively simple, which can be classified into two types, unsigned \emph{int} and signed \emph{int}, with type identifiers \emph{uint} and \emph{int}, respectively. The length of the variable directly determines the range of variable expression. The minimum length of the variable is 8 bits, and the maximum length is 256 bits. For example, \emph{uint8} can represent all natural Numbers between 0 and $2^{8}-1$; \emph{int16} can represent all integers between $-2^{15}$ and $2^{15}-1$.

\subsection{Control Flow Graph}\label{CFG}
Traditionally, the CFG of a program $P$ is a directed graph, in which each node represents a block of code and each edge represents the control flow between blocks~\cite{b20}. In general, the CFG can be represented by a quaternion $G = ( N , E , s , e )$~\cite{b7}, where $N$ represents the collection of program statement nodes; $E$ is the set of directed edges where each edge connecting two nodes is used to indicate the program statement execution sequence; $s$ represents the unique entry (start) node in the CFG; $e$ represents the unique exit (end) node. In addition, there are three main control structures in CFG: \emph{sequence}, \emph{selection}, and \emph{looping} structure. Based on the principle and structure of classical CFG, we realize the construction of CFG that meet the characteristics of \emph{Solidity}.


\subsection{Genetic Algorithm}\label{GA}
GA~\cite{b10} is an optimization algorithm for searching the optimal solution in the problem spaces. In the traditional research on test case generation, GA is used to realize process optimization. Many works show that GA has a good effect in guiding test case generation~\cite{b9,b16,b23}.

In GA, there are three main factors. The first one is the chromosome, which is a specific representation of the target solution of GA. Second, a fitness function is necessary to calculate the fitness value of the individual and the probability of the individual being selected. In addition, three genetic operators, including selection~\cite{b24}, crossover and mutation, are used to generate a new population. The main steps of GA can be summarized as follows:
\begin{enumerate}
\item First-generation population is initialized based on input variable information.
\item The step is further divided in three sub-steps. First, the $dup$ coverage of each individual in the current population is obtained by executing the program after instrumentation. Second, the fitness value of individuals in the population is calculated according to the defined fitness function. Finally, the selection, crossover and mutation operation are carried out to generate a new population.
\item Whether GA meet the termination condition is judged. If the condition meets, the algorithm will output the population with the maximum fitness value. Otherwise, the new population is taken as the current population and return 2).
\end{enumerate}

\section{The ADF-GA Approach}\label{sec approach}

\subsection{Overview of ADF-GA}\label{overview}
The ADF-GA generates valid test cases based on the source codes of smart contracts. The main architecture is shown in Fig.~\ref{fig3}, which contains the following three steps:

\textbf{\emph{Step 1: CFG Construction}}. This step is used to obtain the CFG of a smart contract as an intermediate representation based on source codes.

\textbf{\emph{Step 2: Data Flow Analysis}}. In this step, the data flow analysis method is adopted to obtain the test targets and some input information of GA, including the variable information, the \emph{require} statements and $dup$s in the program.

\textbf{\emph{Step 3: Test Case Generation}}. In this step, we use GA to generate test cases, and the application of program instrumentation to obtain the coverage results and an improved fitness function is also used to calculate individual fitness values.

\subsection{CFG Construction}\label{CC}
\begin{figure}[t]
\centerline{\includegraphics[width=3.8cm,height=6.5cm]{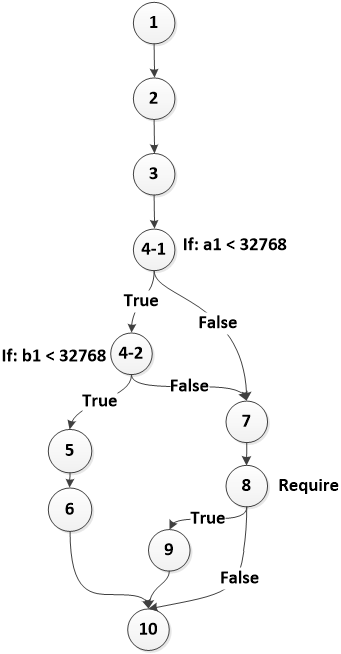}}
\caption{CFG of the sample program in Fig.~\ref{fig1}}
\label{fig4}
\end{figure}

To realize test case generation based on data flow criteria, a reasonable CFG should be constructed first. The characteristics of the CFG are described as follows: 1) It retains the three constructs of the CFG in the traditional languages: \emph{sequence}, \emph{selection}, and \emph{looping} structure. 2) We specify that each node in the CFG represents only a valid statement, not a block. Based on these, the rules for constructing CFG are:
\begin{enumerate}
 \item For every function structure in the contract, a separate CFG (called sub-CFG) is constructed. The \emph{require} statements are processed as \emph{selection} structure. For the \emph{selection} structure generated by a \emph{require} statement, an \emph{arc} is generated and the node of the \emph{require} statement is the \emph{arc} tail. if the condition in the \emph{require} statement is true, the next statement node is the \emph{arc} head. Otherwise, the end node of the sub-CFG is the \emph{arc} head.
  \item We generate an \emph{arc} to represent a function call in the program. The node where the function call occurs is the \emph{arc} tail and the start node of the sub-CFG of the called function is the \emph{arc} head. We then generate another \emph{arc} to represent the return of a function call. The end node of the sub-CFG of the called function is the \emph{arc} tail and the node where the function call occurs is the \emph{arc} head.
\end{enumerate}
According to the rules mentioned above, the CFG constructed for the program segment in Fig.~\ref{fig1} is shown in Fig.~\ref{fig4}.

\subsection{Data Flow Analysis}\label{DFA}
\begin{table}[t]
\caption{Variable Information}
\begin{center}
\begin{tabular}{|c|c|c|}
\hline
\textbf{Variable Name} & \textbf{Variable Type} & \textbf{Variable Length}\\
\hline
 a1 & 0 & 16  \\
 b1 & 0 & 16 \\
 a2 & 0 & 32  \\
 b2 & 0 & 32 \\
 sum1 & 0 & 16 \\
\hline
\end{tabular}
\label{t1}
\end{center}
\end{table}

For the purpose of obtaining the test targets (the collection of $dup$s) and the initial input of GA, we perform the following three phases to achieve a complete data flow analysis.

\quad\emph{a) Variable Information Extraction:} The main purpose of this phase is to get the information of variables existing in the program, which includes variable name, variable type and variable length. Then a list $L_v$ is created based on the information, where we mark the variable type of \emph{uint} as 0, the variable type of \emph{int} as 1, and the length of the variable is the length indicated by its type identifier.

For example, the results of variable information extraction of the code segment shown in Fig.~\ref{fig1} are shown in Table~\ref{t1}.

\quad\emph{b) Require Statements Recognition:} To separately count \emph{require} statements related $dup$s, data flow analysis is performed. ADF-GA identifies the \emph{require} statements. Here we refer to the $dup$s generated by the use of \emph{require} statements and the $dup$s dependent on \emph{require} statements as \emph{require} statement related $dup$s.

For the CFG generated in the previous sub-step, each node is traversed to obtain the location of the node where a \emph{require} statement is located. Table~\ref{t2} is the result of \emph{require} statements identification in the code segment of Fig.~\ref{fig1}.

\begin{table}[t]
\caption{Information of Require Statement}
\begin{center}
\begin{tabular}{|c|c|}
\hline
\textbf{\emph{Require} Statement} & \textbf{Node number} \\
\hline
  require(a2 + b2 \textless= 65535); & 8  \\
\hline
\end{tabular}
\label{t2}
\end{center}
\end{table}

\begin{table}[t]
\caption{Statistical Results of Dup}
\begin{center}
\begin{tabular}{|c|c|c|c|c|}
\hline
\textbf{Variable Name} & \textbf{Def\_node} & \textbf{Use\_node} & \textbf{\emph{N\_dup}} & \textbf{\emph{R\_dup}}\\
\hline
 \multirow{2}{*}{a1} & \multirow{2}{*}{1} & 2,4 & (a1,1,2),(a1,1,4) & \multirow{2}{*}{(a1,1,9)} \\
    &   & 5,9 & (a1,1,5),(a1,1,9) &  \\
\hline
 \multirow{2}{*}{b1} & \multirow{2}{*}{1} & 3,4 & (b1,1,3),(b1,1,4) & \multirow{2}{*}{(b1,1,9)} \\
    &   & 5,9 & (b1,1,5),(b1,1,9) &  \\
\hline
 a2 & 2 & 8 & (a2,2,8) & (a2,2,8) \\
\hline
 b2 & 3 & 8 & (b2,3,8) & (b2,3,8) \\
\hline
 sum & 5,9 & -- & -- & -- \\
\hline
\end{tabular}
\label{t3}
\end{center}
\end{table}

\quad\emph{c) $Dup$ Calculation:} To obtain the set of test targets, we count the number of two types of $dup$s: the first one represents all $dup$s in the program to be tested (abbreviated as $N\_dup$) and the second one represents \emph{require} statements related $dup$s (abbreviated as $R\_dup$). First, two lists $L_{v-d}$ and $L_{v-u}$ are initialized respectively for each variable in $L_v$. $L_{v-d}$ is used to store the definition node of the variable and $L_{v-u}$ is used to store the use node of the variable. Then, the following two stages are performed.

\emph{The first stage:} The generated CFG is traversed in a pre-order way in \textbf{Step 1} to calculate $N\_dup$. The specific operations are as follows:
\begin{enumerate}
\item The CFG is traversed from the start node. If the current traversed node contains a definition of a variable \emph{var1} in $L_v$, the node is added to $L_{v-d}$ of \emph{var1}. If the currently traversed node contains the use of a variable \emph{var2} in $L_v$, the node (denoted as \emph{use}) is added to $L_{v-u}$ of \emph{var2}.
Then the definition node of \emph{var2} (denoted as \emph{def}) is found in the current path to form a $N\_dup$ and put it into the set of test targets, which is represented by $(v,\emph{def},\emph{use})$.
\item The next path in CFG is iterated until all paths are traversed and the $dup$s are constructed in the current traversed path as we performed in 1).
\end{enumerate}

\emph{The second stage:} Counting $R\_dup$. Based on the recognition result of \emph{require} statements, $R\_dup$ is separately counted based on the aforementioned $N\_dup$ set.
Table~\ref{t3} lists the results corresponding to the CFG in Fig.~\ref{fig4}.

\subsection{Test Case Generation Based on Genetic Algorithm}\label{TCGBGA}
ADF-GA adopts GA to guide test case generation. First, we set the termination condition as follows: when the fitness value reaches an optimal value and becomes stable; or the maximum fitness value appears in the historical population and is no longer updated with the execution of the algorithm. Then, we introduce the realization of three factors of GA in ADF-GA.

\begin{figure}
\centering
\subfigure[The encoding of variables of type \emph{uint}]{
   \includegraphics[width=5.7cm,height=1.2cm]{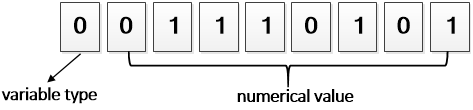}
   }
   \hspace{0in}
\subfigure[The encoding of variables of type \emph{int}]{
   \includegraphics[width=5.7cm,height=1.2cm]{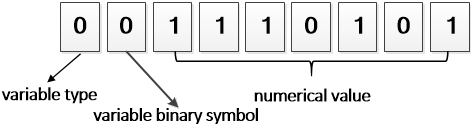}
   }
\caption{Sample of chromosome encoding} \label{fig5}
\end{figure}

\paragraph{Chromosome Coding} To avoid the long encoding of the chromosome due to too many variables contained in the test case and the length of each variable is large, we proposed the concept of sub-chromosome based on the principle of binary coding. One chromosome is a set of sub-chromosomes, and each sub-chromosome is a binary string of one variable. The length of the string of each sub-chromosome is the length of the variable plus 1, where the first bit represents the type of the variable (0 for \emph{uint}, 1 for \emph{int}), and the remaining bits are the binary representation of the decimal value of the variable. Thus, a variable of type \emph{uint8} with a value of 117 can be represented as Fig.~\ref{fig5}. a; a variable of type \emph{int8} with a value of 117 can be represented as Fig.~\ref{fig5}. b.

\paragraph{Fitness Function} Fitness function is the key to guide GA to select good individuals.

The traditional fitness function~\cite{b17} in the GA is shown in~\eqref{eq1}, where $M$ represents the number of $dup$s covered by the current test case and $N$ represents the total number of $dup$s in the program counted in the data flow analysis stage.
\begin{equation}\label{eq1}
fit_{i}=\frac{M}{N}
 \end{equation}

If the tested smart contract contains \emph{require} statements and the generated test cases do not meet the execution conditions of any of these \emph{require} statements, the $dup$ covered is limited and the fitness values of test cases is small. If the majority of the test cases generated are in this case, the algorithm will end up with poor coverage. To alleviate this problem effectively, our approach adjusts the weight of $R\_dup$ when fitness function is designed, so as to pay more attention to the execution of \emph{require} statements in the process of test case selection. The fitness function is expressed as~\eqref{eq2}.
\begin{equation}\label{eq2}
fit_{i}=\frac{(n-m)+(1+\varepsilon)m}{N}
 \end{equation}
where $N$ represents the number of $N\_dup$ contained in the program, $n$ represents the number of $N\_dup$ covered by the current test case, $m$ represents the number of $R\_dup$ covered by the current test case, $\varepsilon$ is the weighted parameter whose value is determined by experiment. According to~\eqref{eq2}, when the parameter takes a appropriate value, the fitness value of individuals with more $R\_dup$ in the population will increase and the probability of being selected as the parent population will increase. Meanwhile, an appropriate parameter also guarantees that those individuals with most $N\_dup$ coverage still have a large probability of being selected. Therefore, the population can be oriented towards achieving the overall better results.

\paragraph{Genetic Operators} Using GA to generate a new population based on the current population requires the following three operations: selection, crossover, mutation.

\textbf{Selection}: Our approach adopts the roulette wheel selection algorithm~\cite{b24}, whose basic idea is to calculate the probability of each individual being selected according to the fitness value. The specific operations are as follows:
\begin{enumerate}
\item For the population of size $n$, the fitness value of each individual in the population is calculated according to \eqref{eq2}, denoted as $f_{i}$, $i=1 \cdots n$. Then the probability of the individual being selected is further calculated based on \eqref{eq3}, denoted as $p_{i}$, $i=1 \cdots n$.
\item The cumulative probability of each individual is calculated according to \eqref{eq4}, denoted as $p_{i}'$, $i=1 \cdots n$, where $p_{1}'= p_{1}$, $p_{n}' = 1$.
\item A number $r$ is randomly generated between 0 and 1. The selected individual is determined according to the numerical value $r$ and the cumulative probability $p'$. If $p_{i-1}'<r\leq p_{i}'$, the $i^{th}$ individual is selected.
\end{enumerate}

\begin{equation}\label{eq3}
p_{i}=\frac{f_{i}}{\sum_{i=1}^{n}f_{i}}
 \end{equation}

\begin{equation}\label{eq4}
p_{i}'=\sum_{j=0}^{i}p_{j}
 \end{equation}
\quad It can be found from \eqref{eq3} that the greater the fitness value of the individual, the greater the probability of being selected.

\begin{figure}[t]
\centering
\includegraphics[width=8cm,height=1cm]{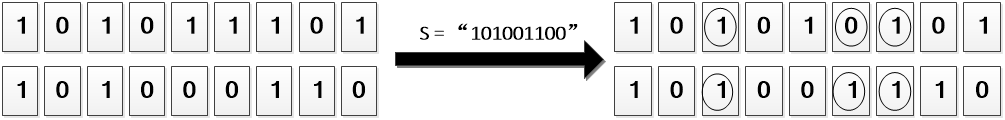}
\caption{Sample of sub-chromosome crossing}\label{fig6}
\end{figure}

\textbf{Crossover}: Uniform crossover is used. The basic operation is to perform crossover on each sub-chromosome within two paired individuals and each gene in the two paired sub-chromosomes is exchanged with equal probability. The details are described as follows: if the two paired sub-chromosomes are expressed as $X=x_1 x_2 \cdots x_m$ and $Y=y_1 y_2 \cdots y_m$ respectively. Before the crossover is performed, a binary string $S$ of the same length as the two sub-chromosomes is generated randomly, denoted by $S=s_1 s_2 \cdots s_m$. According to the chromosome coding, $s_1$ represents the type of variable, so the crossover is done bit by bit from $s_2$ to $s_m$. If the value of $s_i$ is 0, the value of $x_i$ and $y_i$ are maintained. If it is 1, the value of the $x_i$ and $y_i$ is exchanged. Fig.~\ref{fig6} shows a sub-chromosome crossing process.

\textbf{Mutation}: The goal of the mutation operation is to mutate one or more of the genes in a chromosome string, which is expressed as changing the original code of 0 to 1 and the original code of 1 to 0 in ADF-GA. Specifically, the mutation probability $P_m$ is set in advance. For each chromosome in the population, each of its sub-chromosomes is processed one by one. For each encoding gene on the sub-chromosome, a random number $r$ between 0 and 1 is generated. When $r<P_m$, the mutation operation is performed.

\section{Experimental Evaluation}\label{sec_validation}

\subsection{Experimental Setup}\label{ES}

\begin{table}[t]
\caption{Information About the Contract Programs}
\begin{center}
\begin{tabular}{|c|c|c|c|c|}
\hline
\multirow{2}{*}{\textbf{Program}} & \multirow{2}{*}{\textbf{LOC}} & \textbf{Number of} & \textbf{Number of} & \textbf{Number of} \\
   & & \textbf{Function} & \textbf{\emph{N\_dup}} & \textbf{\emph{R\_dup}} \\
\hline
 getcenter & 47 & 1 & 28 & 0 \\
 getsum & 55 & 1 & 23  & 0 \\
 safe\_add & 22 & 1 & 10 & 4 \\
 math\_op & 84 & 7 & 102 & 0 \\
 safe\_buy & 88 & 5 & 29 & 12 \\
 fundraise & 123 & 4 & 33 & 14 \\
 math\_op\_reqiure & 126 & 7 & 108 & 22 \\
 operationofarray & 155 & 11 & 62 & 25  \\
 info\_manage\_sys & 163 & 5 & 74 & 30  \\
 trade & 202 & 10 & 64 & 30 \\
 geometry & 246 & 9 & 93 & 50  \\
\hline
\end{tabular}
\label{t4}
\end{center}
\end{table}

\quad\emph{a) Experiment Environment:} We conduct our experiments in a computer system with Intel(R) Core(TM) i5-8300H CPU @2.30GHz, 8.00GBRAM, Windows 10. MATLAB R2016a is used to build the model for GA, and Remix 0.4.22 is used to execute the \emph{Solidity} smart contracts.


\quad\emph{b) Solidity programs:} We consider the following two factors when selecting the representative \emph{Solidity} programs: 1) the experimental cost caused by the limitation of the smart contract operating environment; 2) diversity and universality of the experimental programs. Therefore, we choose the following 11 smart contracts to validate the effectiveness of ADF-GA. These contract programs basically contain various structural statements commonly used in \emph{Solidity} and have different code size as well as the numbers of functions and $dups$.
The specific information of them is shown in Table~\ref{t4}, where the first column represents the number of valid lines of code that contain only partial comments; column 4 represents the number of $N\_dup$ contained; column 5 represents the number of $R\_dup$ contained. These smart contracts are new programs that have been artificially modified and populated to perform different functions based on a publicly available smart contract data set\footnote{https://github.com/fictional-tribble-2/not-so-smart-contracts} on GitHub. We have compiled and run these programs to verify their availability and uploaded them to GitHub\footnote{https://github.com/MoiraYjn/test-demo-}.

\quad\emph{c) Comparative Approaches:} At present, there is no other work on the generation of  data flow test cases for smart contracts. Therefore, we first implement ADF-GA and then based on smart contracts we re-implement other two representative approaches: random testing approach~\cite{b26} and the approach proposed in~\cite{b17}, which originally designed for \emph{C\#}. In the following, we refer these two approaches as RT and GA-C\#.

\subsection{Experimental Results}\label{ER}

\quad\emph{a) Parameters:} To obtain a better value of parameter $\varepsilon$, we conduct a set of experiments based on four contracts. Based on the same first-generation population, we change the value of parameter $\varepsilon$ to perform multiple experiments and determine the final value of $\varepsilon$ by comparing the coverage of test cases generated at different values of $\varepsilon$ to $N\_dup$ and $R\_dup$.

\begin{figure}[t]
\centering
\subfigure[]{
   \includegraphics[width=8.5cm,height=3.5cm]{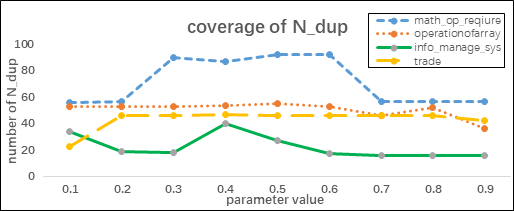}
   }
   \hspace{0in}
\subfigure[]{
   \includegraphics[width=8.5cm,height=3.5cm]{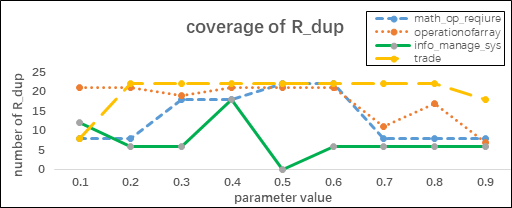}
   }
\caption{Coverage of $N\_dup$ and $R\_dup$ with different parameters-1} \label{fig7}
\end{figure}

First, we take 0.1 as the step size.
Then, we count the number of $N\_dup$ and $R\_dup$ covered by the optimal test cases generated by ADF-GA at different parameter, respectively. The experimental results are shown in Fig.~\ref{fig7}, in which, (a) shows the coverage of the $N\_dup$ when the parameter takes different values; (b) shows the coverage of the $R\_dup$.

\begin{figure}[t]
\centering
\subfigure[]{
   \includegraphics[width=8.5cm,height=3.5cm]{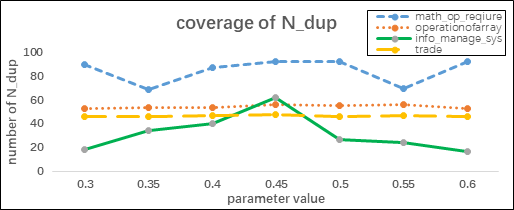}
   }
   \hspace{0in}
\subfigure[]{
   \includegraphics[width=8.5cm,height=3.5cm]{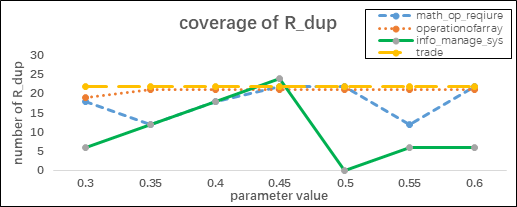}
   }
\caption{Coverage of $N\_dup$ and $R\_dup$ with different parameters-2} \label{fig8}
\end{figure}

\begin{table*}[t]
\caption{Iterations in Genetic Algorithm}
\begin{center}
\begin{tabular}{|l|c|c|c|c|c|}
\hline
   \multirow{3}*{\textbf{Program}} & \multicolumn{2}{c|}{\multirow{2}*{\textbf{Total  iterations}}} & \multicolumn{2}{c|}{\textbf{Iterations when the maximum fitness value}} & \textbf{Iterations Of ADF-GA} \\
   & \multicolumn{2}{c|}{} & \multicolumn{2}{c|}{\textbf{is reached for the first time}} & \textbf{to achieves the best} \\
   \cline{2-5}
   & \textbf{ADF-GA} & \textbf{GA-C\#} & \textbf{ADF-GA} & \textbf{GA-C\#} & {\textbf{coverage Of GA-C\#}} \\
\hline
  \textbf{getcenter} & 8 & 9 & 2 & 3 & 2  \\
  \textbf{getsum} & 21 & 21 & 8 & 9 & 8  \\
  \textbf{safe\_add} & 7 & 7 & 1 & 1 & 1  \\
  \textbf{math\_op} & 30 & 29 & 10 & 11 & 10 \\
  \textbf{safe\_buy} & 25 & 20 & 16 & 3 & 2  \\
  \textbf{fundraise} & 23 & 14 & 9 & 3 & 2  \\
  \textbf{math\_op\_reqiure} & 23 & 32 & 17 & 25 & 14 \\
  \textbf{operationofarray} & 19 & 28 & 14 & 18 & 12  \\
  \textbf{info\_manage\_sys} & 27 & 27 & 19 & 19 & 18  \\
  \textbf{trade} & 32 & 24 & 15 & 18 & 3  \\
  \textbf{trade} & 18 & 31 & 10 & 15 & 10  \\
  \textbf{geometry} & 21.18 & 22 & 11 & 11.36 & 7.45 \\
\hline
\end{tabular}
\label{t5}
\end{center}
\end{table*}
It can be seen from Fig.~\ref{fig7}. a, when the value of the parameter is between 0.3 and 0.6, a better $N\_dup$ coverage can be achieved. The experimental results in Fig.~\ref{fig7}. b also show that when parameter $\varepsilon$ is between 0.3 and 0.6, the test cases can cover most $R\_dup$s.

To obtain a more accurate parameter, we set the value between 0.3 and 0.6 with the step size of 0.05 for another set of experiments. The parameter $\varepsilon$ is 0.3, 0.35, 0.4, 0.45, 0.5, 0.55, and 0.6, respectively. Fig.~\ref{fig8} show the experimental results. It is clear that there is a difference in the number of covered $dup$ ($R\_dup$ and $N\_dup$) at different parameters. Both the number of covered $R\_dup$ and the number of covered $N\_dup$ are the best when $\varepsilon$ is 0.45. Consequently, 0.45 is chosen as the best experimental parameters.

\begin{figure}[t]
\centering
\subfigure[]{
   \includegraphics[width=8.5cm,height=4cm]{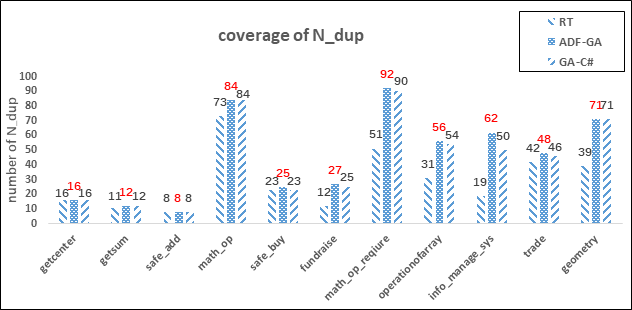}
   }
   \hspace{0in}
\subfigure[]{
   \includegraphics[width=8.5cm,height=4cm]{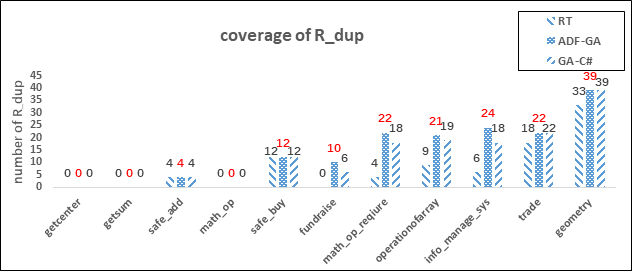}
   }
\caption{Coverage of $N\_dup$ and $R\_dup$ with different approaches}
\label{fig9}
\end{figure}

\quad\emph{b) Coverage :} To validate whether ADF-GA can achieve higher coverage of $dup$, we conduct a set of dedicated experiments on the 11 smart contracts using RT, GA-C\# and ADF-GA, respectively, and compared the experimental results.

Fig.~\ref{fig9} shows a comparison of the coverage of test cases generated by three approaches. First, compared to RT, it is obvious that ADF-GA can cover more test targets ($dup$), This is because ADF-GA uses genetic algorithm that can effectively guide test case generation. Furthermore, it can be seen that for all the tested contract programs, ADF-GA covers more or the same number of $N\_dup$ and $R\_dup$ compared with GA-C\#. The reason is that the fitness function of ADF-GA can further differentiate two kinds of $dups$, which makes the coverage of $R\_dup$ more sensitive to the generated test cases and ultimately increases the coverage.

\quad\emph{c) Performance:} To validate whether ADF-GA can effectively guide the execution of the algorithm, we apply ADF-GA and GA-C\# to perform test case generation and analyze experimental results by comparing the number of the iterations in the genetic algorithms. 

We respectively count the total iterations in GA, the number of iterations when the maximum fitness value is reached for the first time during the execution of GA, and the iterations when ADF-GA achieves the best coverage that GA-C\# can obtain. The results are shown in Table~\ref{t5}.
First, it can be seen that ADF-GA has fewer total iterations for some contracts and for some other contracts GA-C\# has fewer total iterations.
In general. we can draw the conclusion that the total number of iterations in GA in ADF-GA is slightly fewer than that in GA-C\#.
In addition, comparing the iterations when the first maximum fitness value is reached for the first time, ADF-GA requires slightly fewer iterations than GA-C\#, which iterates 11.36 times, while ADF-GA iterates 11 times. Finally, we record the iterations that ADF-GA needs to achieve the best coverage of GA-C\# and then compare it with the iterations that GA-C\# needs to achieve the best coverage for the first time. The results show that it takes only 7.45 iterations for ADF-GA. Therefore, we can get the conclusion that ADF-GA can effectively guide the generation of test cases.
In summary, ADF-GA can effectively generate test cases covering more test targets.

\section{Conclusions and future work}\label{sec_conclusions}
This paper presents a novel approach called ADF-GA for the all-uses data flow criterion based test case generation of smart contracts, which uses genetic algorithm to realize the dynamic test of smart contracts. 
We compare ADF-GA with two traditional approaches and the results show that ADF-GA can generate available test cases more efficiently.

For future work, first, we need to further optimize the selection and mutation operation to generated test cases. Second, we will explore better fitness functions to evaluate the generated test cases. Finally, due to the limitations of the execution environment of smart contracts, we will try to solve the constraint of the environment on dynamic testing of smart contract programs.

\section*{Acknowledgment}
The work is supported by the National Natural Science Foundation of China under Grant No. 61702159 and No. 61572171, the Natural Science Foundation of Jiangsu Province under Grant No. BK20170893 and No. BK20191297.


\vspace{12pt}

\end{document}